\documentclass[fleqn,twoside,twocolumn,nofootinbib,showkeys]{revtex4} %
\usepackage[nocpr,nopacs]{ujp_UTF8} 

\begin{document}
\title[Characteristics of the Invariant
Measure of the Strange Attractor]
{CHARACTERISTICS OF THE INVARIANT\\ MEASURE OF THE STRANGE ATTRACTOR\\ OF THE BACTERIA MATHEMATICAL MODEL}%
\author{V.~Grytsay}
\affiliation{\bitp}
\address{\bitpaddr}
\email{vgrytsay@bitp.kiev.ua}

\udk{577.3}  \razd{\seciii}

\autorcol{V.I.\hspace*{0.7mm}Grytsay}

\setcounter{page}{1}%

\begin{abstract}
The bacteria metabolic process of open nonlinear dissipative system
far from equilibrium point is modeled using classical methods of
synergetics.\,\,The invariant measure and its convergence in the
phase space of the system was obtained in strange attractor
mode.\,\,The distribution of point density of trajectory
intersection of phase space cells with maximum invariant measure and
convergence in time of its average value was obtained.\,\,The result
concluded is that the value of an invariant measure can be a
characteristic of the transitional process of adaptation of cell
metabolic process to change outside environment.
\end{abstract}
 \maketitle

\section{Introduction}

This article investigates the mathematical model of bacteria
metabolic process constructed in \cite{1,2,3}. The experimental
results of biochemical process of \textit{Arthrobacter globiformis}
cells in bioreactor \cite{4,5} were used as a base for construction
of mathematical model.\,\,The model shows the main metabolic
connections of oxygen-breathing bacteria.\,\,The metabolic process
in a cell was considered as an open dissipative system.\,\,The
system has two main self organized subsystems of the dissipative
system:
substrate transformation and a breath chain.

\section{Mathematical Model}

\begin{figure}%
\vskip1mm
\includegraphics[width=6cm]{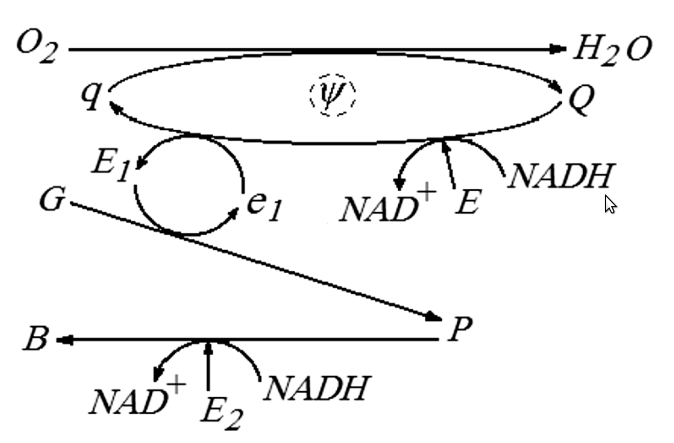}
\vskip-3mm\caption{The main scheme of a cell metabolic process
}\label{fig:1}\vspace*{2mm}
\end{figure}

The mathematical model was constructed according to the general
scheme of cell metabolic process Fig.~\ref{fig:1} and was described
in the system (\ref{1})--(\ref{10})
\cite{6,7,8,9,10}:
\begin{equation}\label{1}
\frac{dG}{dt}=\frac{G_0}{N_3+G+\gamma_2\psi}-l_1V(E_1)V(G)-\alpha_3G,
\end{equation}\vspace*{-5mm}
\begin{equation}\label{2}
\frac{dP}{dt}=l_1V(E_1)V(G)-l_2V(E_2)V(N)V(P) -\alpha_4P,
\end{equation}\vspace*{-6mm}
\begin{equation}\label{3}
\frac{dB}{dt}=l_2V(E_2)V(N)V(P)- k_1V(\psi)V(B) -\alpha_5B,
\end{equation}
\[
\frac{dE_1}{dt}=E_{10}\frac{G^2}{\beta_1+G^2}\left(\!1-\frac{P+mN}{N_1+P+mN}\!\right)-
\]\vspace*{-7mm}
\begin{equation}\label{4}
-\,l_1V(E_1)V(G)+l_4V(e_1)V(Q) -a_1E_1,
\end{equation}\vspace*{-7mm}
\begin{equation}\label{5}
\frac{de_1}{dt}=-l_4V(e_1)V(Q)+l_1V(E_1)V(G) -\alpha_1e_1,
\end{equation}\vspace*{-7mm}
\[
\frac{dQ}{dt}=6lV(2-Q)V(O_2)V^{(1)}(\psi)-l_6V(e_1)V(Q)\,-
\]\vspace*{-7mm}
\begin{equation}\label{6}
-\,l_7V(Q)V(N),
\end{equation}\vspace*{-7mm}
\begin{equation}\label{7}
\frac{dO_2}{dt}=\frac{O_{20}}{N_5+O_2}-lV(2-Q)V(O_2)V^{(1)}(\psi)-\alpha_7O_2,
\end{equation}\vspace*{-10mm}
\[
\frac{dE_2}{dt}=E_{20}\frac{P^2}{\beta_2+P^2}\frac{N}{\beta+N}\left(\!1-\frac{B}{N_2+B}\!\right)-
\]\vspace*{-7mm}
\begin{equation}\label{8}
-\,l_{10}V(E_2)V(N)V(P) -\alpha_2E_2,
\end{equation}\vspace*{-7mm}
\[
\frac{dN}{dt}=-l_2V(E_2)V(N)V(P)-l_7V(Q)V(N)\,+
\]\vspace*{-7mm}
\begin{equation}\label{9}
+\, k_2V(B)\frac{\psi}{K_{10}+\psi}+\frac{N_0}{N_4+N} -\alpha_6N,
\end{equation}\vspace*{-7mm}
\begin{equation}\label{10}
\frac{d\psi}{dt}=l_5V(E_1)V(G)+l_8V(N)V(Q) -\alpha\psi.
\end{equation}
where $V(X)=X/(1+X)$, $V^{(1)}(\psi )=1/(1+\psi ^2)$.

The parameters of the model are $l=l_1 =k_1 =0.2$; $l_2 =l_{10}
=0.27$; $l_5 =0.6$; $l_4 =l_6 =0.5$; $l_7 =1.2$; $l_8 =2.4$; $k_2
=1.5$; $E_{10} =3$; $\beta _1 =2$; $N_1 =0.03$; $m=2.5$; $\alpha
=0.033$; $a_1 =0.007$; $\alpha _1 =0.0068$; $E_{20} =1.2$; $\beta
=0.01$; $\beta _2 =1$; $N_2 =0.03$; $\alpha _2 =0.02$; $G_0 =0.019$;
$N_3 =2$; $\gamma _2 =0.2$; $\alpha _5 =0.014$; $\alpha _3 =\alpha
_4 =\alpha _6 =\alpha _7 =0.001$; $O_{20} =0.015$; $N_5 =0.1$; $N_0
=0.003$; $N_4 =1$; $K_{10} =0.7$.

The nonlinear differential system (\ref{1})--(\ref{10}) was solved
using Runge--Kutta--Metson method with accuracy $10^{-12}$.

The study of the solutions of the mathematical model
(\ref{1})--(\ref{10}) was performed using nonlinear differential
equation theory \cite{11,12,13,14} and developed methods of
mathematical modeling of biochemical systems by the author and other
researchers
\cite{15,16,17,18,19,20,21,22,23,24,25,26,27,28,29,30,31,32,33,34,35,36,37,38,39,40,41,42,43}.

Using this model, all possible modes of the me\-ta\-bolic process as
a function of small parameter were investigated.\,\,Self
organization modes and dynamical chaos were found
\cite{28,29,30}.\,\,Lia\-pu\-nov exponents were obtained.\,\,A
spectral analysis of the system solutions was carried out.\,\,The
stability of the modes dynamic were studied.

\section{Results of Studies}

In the work, we investigate experimental modes of the cell metabolic
process that may arise in bioreactor.\,\,In previous papers, we
investigated and described kinetic curves of self organization modes
in detail.\,\,Strange attractors modes can not be described by
kinetic curves.\,\,It is because calculation and experimental
characteristics are not comparable because of exponential trajectory
run off and hypersensitivity of the system to initial data.

The author suggests to describe such types of modes of the system by
calculating invariant measure.\,\,It defines probability of existing
trajectory in different region of the phase space.\,\,The work
continues investigation of invariant measures for strange attractors
of the systems started in work~\cite{37}.

From Krylov--Bogolyubov theorem, in a case continuous mapping and
compact phase space of dynamical system (\ref{1})--(\ref{10}), there
exists at least one invariant measure $\mu_i$
\cite{11}.\,\,Ob\-tained phase portraits and invariant measures
confirmed that the system complies with these requirements.

Let us investigate strange attractor mode of the system $13 \times
2^x$ ($\alpha=0.03217)$.

The strange attractor was created as result a funnel.\,\,In this
area the its trajectories are mixing.

Let us investigate properties of the invariant \mbox{measure.}

\begin{figure*}%
\vskip1mm
\includegraphics[width=15.4cm]{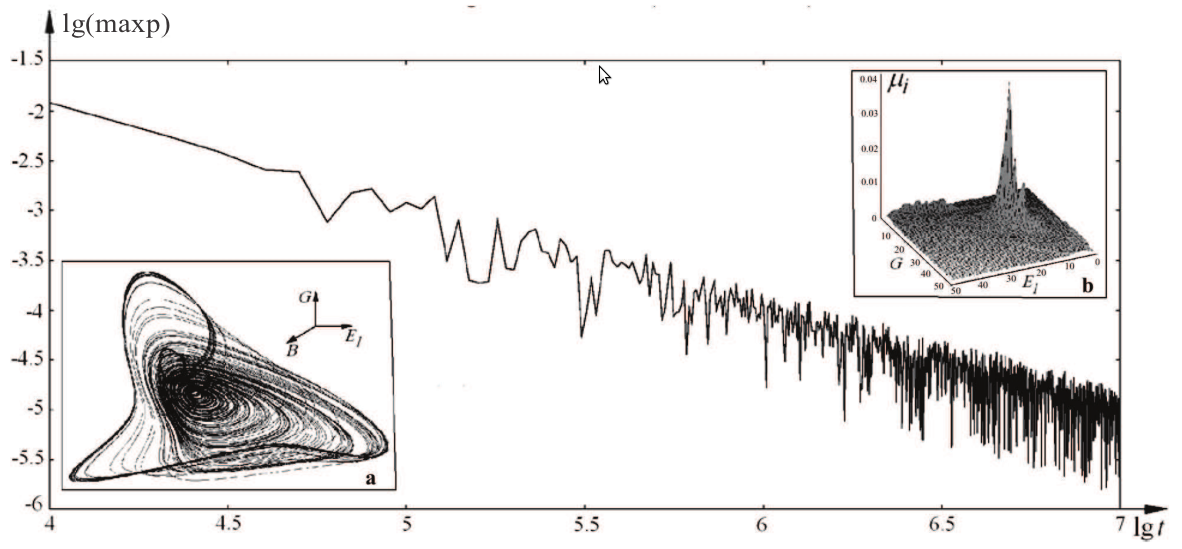}
\vskip-2mm\parbox{15.4cm}{\caption{Graph of convergence of the
invariant measure of the strange attractor for the system $13 \times
2^x$ ($\alpha=0.03217)$, where: \textbf{a} -- a projection of the
phase portrait of the attractor in 3d phase space $E_1, G, B$;
\textbf{b} -- a histogram of the projection of the invariant measure
of the strange attractor
onto the plane $G, E_1$\label{fig:2}}}
\end{figure*}

A convergence graph of invariant measure for the strange attractor
was constructed as shown in Fig.~\ref{fig:2}.\,\,Cal\-cu\-lations
show that changing of amount of mapping points does not influence
the probability of visiting the trajectory of each cell.\,\,Time
shifting along trajectory does not influence the probability
too.\,\,It means an invariance of a measure for the strange
attractor.\,\,The peak of the invariant measure indicates an
attracting set of the strange attractor in the mixing funnel.

Let us obtain attractor convergence $13 \times 2^x$ ($\alpha=$
$=0.03217)$, for each of 10000 iterations.

\begin{figure*}%
\vskip2mm
\includegraphics[width=15.4cm]{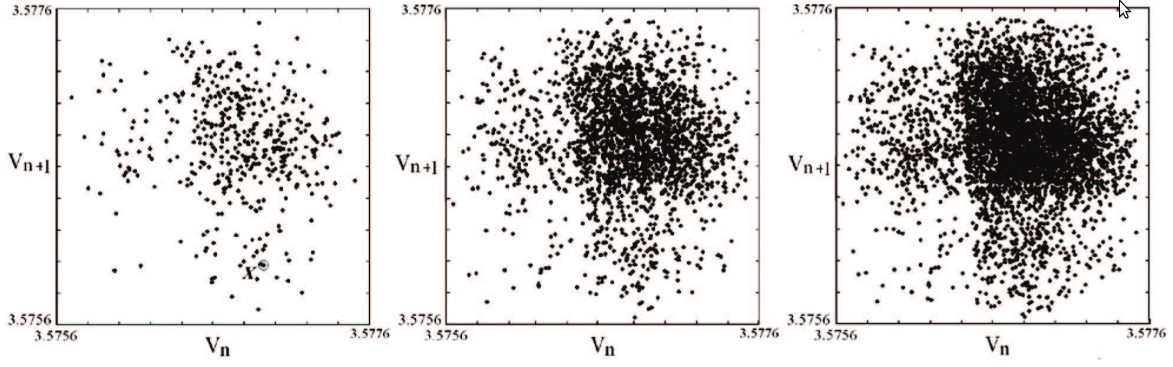}\\
{\it a\hspace{5cm}b\hspace{5cm}c}\\[2mm]
\includegraphics[width=15.4cm]{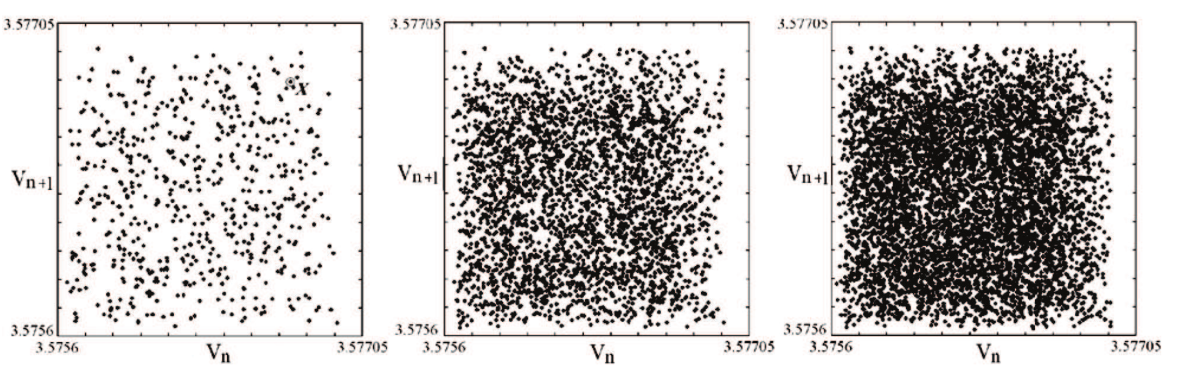}\\
{\it d\hspace{5cm}e\hspace{5cm}f}
\vskip-2mm\parbox{15.4cm}{\caption{The evolution for density points
distribution of intersection strange attractor trajectory $13 \times
2^x$ cells of phase space with maximum invariant measure for
$N=200^{10}$ cells: $a$ ($\sum n =465$, $t=4\times 10^6$), $b$
($\sum n =2298$, $t=2\times 10^7$), $c$ ($\sum n =4592$, $t=4\times
10^7$); for $N=1000^{10}$ cells: $d$ ($\sum n =598$, $t=4\times
10^7$), $e$ ($\sum n =3103$, $t=2\times 10^8$), $f$ ($\sum n =6110$,
$t=4\times 10^8$)
  \label{fig:3}}}
\end{figure*}

\begin{figure}%
\vskip1mm
\includegraphics[width=8.1cm]{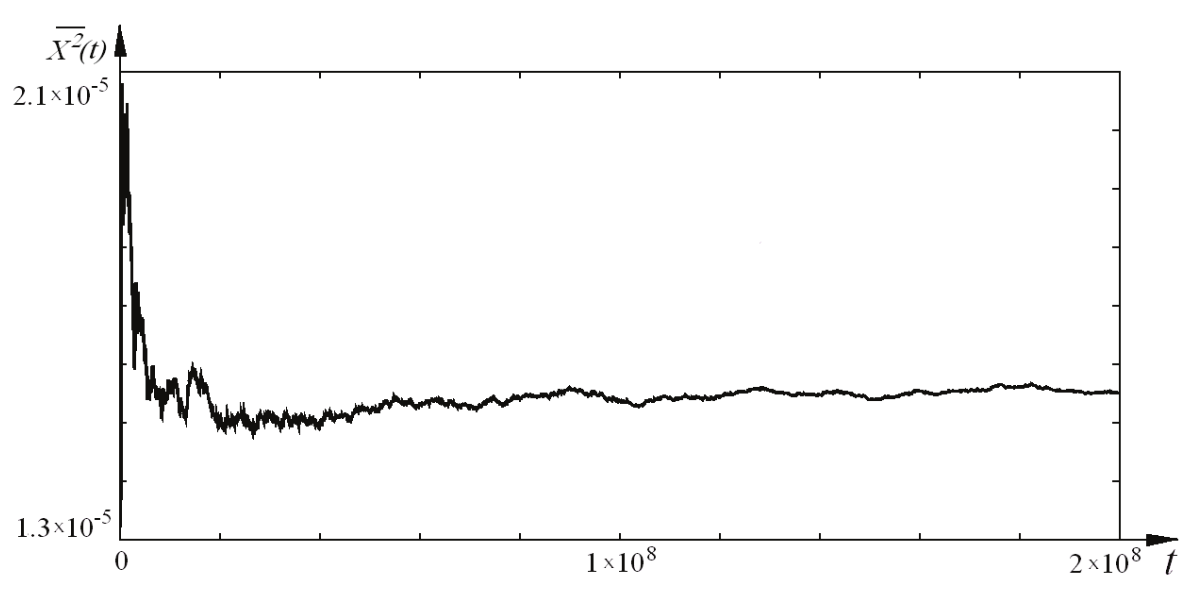}
\vskip-3mm\caption{The graph of convergence of the mean over time
for strange attractor of the system $13 \times 2^x$ in the cell with
maximum invariant measure ($N=1000^{10}$, $\sum n =3103$, $t=2\times
10^8$) }\label{fig:4}
\end{figure}

From the graph of convergence measure it can be seen that the
measure tends to converge to its average value.\,\,The value of the
measure is decreasing~as~$1/t$.\looseness=1

Let us investigate this situation.\,\,Lets obtain a density
distribution of points of intersection of strange attractor
trajectory for cell of phase space with maximum invariant measure.

Let us obtain and compare invariant measure for $N=200^{10}$
(Fig.~\ref{fig:3}, \textit{a, b, c}) and $N=1000^{10}$
(Fig.~\ref{fig:3}, \textit{d, e, f}).

The change with time of distribution density of intersection points
can be seen in Fig.~\ref{fig:3}.\,\,Where $X$ is the first
intersection point of the cell.\,\,In both cases, intersection
points group in compressed areas.

Let us obtain a convergence of the mean over time in the cell with
maximum invariant measure ($N=$ $=1000^{10}$, $\sum n =3103$,
$t=2\times 10^8$) (Fig.~\ref{fig:4}).

It can be concluded from the graph that variation of the mean slowly
decreases with time.\,\,The system stabilizes in a new
auto-oscillating mode.

Obtained invariant measure and its convergence show adaptive
capabilities of metabolism in cell in self organization process to
environment of the dissipative system.\,\,The metabolic process is
maintained by the cell at the vicinity of average level of its
\mbox{metabolites.}\looseness=1

\section{Conclusions}

Strange attractor mode of the cell metabolic process was
investigated by the mathematical model.\,\,The possibility of
application of the calculation of the invariant measure for chaotic
modes of the model has been investigated.\,\,The distribution
density of intersection points of trajectories of the cell in a
phase space correspondent to the maximum invariant measure is
found.\,\,The convergence in time of its average value is
demonstrated.\,\,It is concluded that the value of invariant measure
and its convergence show adaptive capabilities of cell metabolism
during self organization as a response to change in
environment.\,\,Main\-tenance of cell metabolites around their
average values is demonstrated.

\vskip3mm \textit{The work was supported by project
PKNo\,0120U101347 the National Academy of Science\linebreak of
Ukraine.}

\vspace*{-5mm}
\rezume{%
В.Й.\,Грицай}{ОСОБЛИВОСТІ ІНВАРІАНТНОЇ\\ МІРИ ДИВНОГО АТРАКТОРА
МАТЕМАТИЧНОЇ\\ МОДЕЛІ БАКТЕРІЇ} {Використовуючи класичні методи
синергетики, проведено моделювання метаболiчного процесу бактерії~--
відкритої нелінійної дисипативної системи, далекої від рівноваги. В
режимі дивного атрактора розраховується інваріантна міра та її
збіжність у фазовому просторі системи. Розраховано розподіл густини
точок перетину траєкторією комірки фазового простору з максимумом
інваріантної міри та  збіжність по часу її  середнього значення.
Зроблено висновок: величина  інваріантної міри може бути
характеристикою перехідного процесу адаптації метаболізму клітини до
змін у навколишньому середовищі.}{\textit{К\,л\,ю\,ч\,о\,в\,і\,
с\,л\,о\,в\,а}: математична модель, метаболічний процес, дивний
атрактор, фазовий простір, інваріантна міра, збіжність.}


\begin{thebibliography}{99}
\bibitem{1} V.P. Gachok, V.I. Grytsay. The kinetic model of macroporous granule with the regulation of biochemical processes. \textit{Dokl. Akad. Nauk SSSR} \textbf{282}, No.~1, 51 (1985).
\bibitem{2} V.P. Gachok, V.I. Grytsay, A.Yu. Arinbasarova, A.G.~Me\-dentsev, K.A. Koshcheyenko, V.K. Akimenko. Ki\-ne\-tic model of hydrocortisone 1-en-dehydrogenation by \textit{Arthrobacter globiformis}. \textit{Biotechn. Bioengin}. \textbf{33}, 661 (1989).
\bibitem{3} V.P. Gachok, V.I. Grytsay, A.Yu. Arinbasarova, A.G.~Me\-dentsev, K.A. Koshcheyenko, V.K. Akimenko. Ki\-ne\-tic model for the regulation of redox reaction in steroid transformation by \textit{Arthrobacter globiformis} cells. \textit{Biotechn. Bioengin}. \textbf{33}, 668 (1989).
\bibitem{4} A.A. Akhrem, Yu.A. Titov. \textit{Steroids and Microorganisms} (Nauka, 1970) (in Russian).\vspace*{0.2mm}
\bibitem{5} A.G. Dorofeev, M.V. Glagolev, T.F. Bondarenko, N.S.~Pa\-nikov. Unusual growth kinetics of Arthrobacter gtlobiformis and its explanation. \textit{Mikrobiol.} \textbf{61}, 33 (1992).\vspace*{0.2mm}
\bibitem{6} V.I. Grytsay. The self-organization in a macroporrous structure of a gel with immobilized cells. The kinetic model of a bioselective membrane of a biosensor. \textit{Dopov. NaN Ukr.} No.~2, 175 (2000).\vspace*{0.2mm}
\bibitem{7} V.I. Grytsay. The self-organization in a reaction-diffsion porous medium. \textit{Dopov. NaN Ukr.} No.~3, 201 (2000).\vspace*{0.2mm}
\bibitem{8} V.I. Grytsay. Ordered structures in the mathematical model of a biosensor. \textit{Dopov. NAN Ukr.} No.~11, 112 (2000).\vspace*{0.2mm}
\bibitem{9} V.I. Grytsay. Structural instability of a biochemical process. \textit{Ukr. J. Phys.} \textbf{55}, No.~5, 599 (2010).
\bibitem{10} V.I. Grytsay, I.V. Musatenko. Self-oscillatory dynamics of the metabolic process in a cell. \textit{Ukr. Biochem. Zh.} \textbf{85}, No.~2, 93
(2013).
\bibitem{11} N.N. Bogolubov. Collected works in 12 volumes (Nauka,  2005) (in Russian).\vspace*{0.2mm}
\bibitem{12} S.P. Kuznetsov. \textit{Dynamial Chaos} (Fizmatlit,  2001) (in Russin).\vspace*{0.2mm}
\bibitem{13} V.S. Anishchenko. \textit{Complex Oscillations in Simple Systems} (Nauka,  1990) (in Russian).\vspace*{0.2mm}
\bibitem{14} G.G. Malinetskii, A.B. Potapov. Nonlinear dynamics and
chaos: Basic concepts (URSS,  2006) (in Russian).\vspace*{0.2mm}
\bibitem{15} V.I. Grytsay. Ordered and chaotic structures in the reaction-diffuzive porous media \textit{Visn. Kyiv. Univ.} No.~2, 394 (2002).\vspace*{0.2mm}
\bibitem{16} V.I. Grytsay. Condidions of self-organization of prostacyclin and thromboxane. \textit{Visn. Kyiv. Univ.} No.~3, 372\linebreak (2002).\vspace*{0.2mm}
\bibitem{17} V.I. Grytsay, V.P. Gachok. Regimes of self-irganization in system of protacyclin and thromboxan. \textit{Visn. Kiev Univ.} No.~4, 365 (2002).\vspace*{0.2mm}
\bibitem{18} V.I. Grytsay, V.P. Gachok. Ordered structures in mathematical sytem of prostacyclin and tromboxan model. \textit{Visn. Kyiv Univ.} No.~1, 338 (2003).\vspace*{0.2mm}
\bibitem{19} V.I. Grytsay. Processes modeling of the multienzyme prostacyclin and thromboxan sytem. \textit{Visn. Kyev. Univ.} No.~4, 379 (2003).\vspace*{0.2mm}
\bibitem{20} V.V. Andreev, V.I. Grytsay. Modeling of the inactive zones in porous catalyst granules and in a biosensor. \textit{Matem. Modelir.} \textbf{17}, No.~2, 57 (2005).\vspace*{0.2mm}
\bibitem{21} V.V. Andreev, V.I. Grytsay. Influence of diffusion reaction processes non-informaty on structures formation in porous medium. \textit{Matem. Modelir.} \textbf{17}, No.~6, 3 (2005).\vspace*{0.2mm}
\bibitem{22} V.I. Grytsay, V.V. Andreev. The diffusion role on non-active structures formation in porous reaction-diffusion medium. \textit{Matem. Modelir.} \textbf{18}, No.~12, 88 (2006).
\bibitem{23} V.I. Grytsay. Uncertainty in the evolution structure of reactiondiffusion medium of bioreactor. \textit{Biofiz. Visn.} Iss.~2 (19), 92 (2007).
\bibitem{24} V. Grytsay. Unsteady conditions in a porous reaction-diffusion medium. \textit{Romanian J. Biophys.} \textbf{17}, No.~1,  55 (2007).
\bibitem{25} V.I. Grytsay. Morphogenetic field forming and stability of bioreactor immobilization cells. \textit{Biofiz. Visn.} Iss.~1 (20), 48(2008).
\bibitem{26} V.I. Grytsay. Prediction structural instability and type attractor of biochemical process. \textit{Biofiz. Visn.} Iss.~2 (23), 77 (2009).

\bibitem{27} V.I. Grytsay, I.V. Musatenko. Self-organization and fractality in a metabolic process of the Krebs cycle. \textit{Ukr. Biokhim. Zh.} \textbf{85}, No.~5, 191 (2013).

\bibitem{28} V.I. Grytsay, I.V. Musatenko. The structure of a chaos of strange attractors within a mathematical model of the metabolism of a cell. \textit{Ukr. J. Phys.} \textbf{58}, No.~7, 677 (2013).
\bibitem{29} V. Gytsay, I. Musatenko. A mathematical model of the metabolism of a cell. Self-organization and chaos. \textit{Chaotic modeling and Simulation (CMSIM)} No.~4, 539 (2013).
\bibitem{30} V.I. Grytsay, I.V. Musatenko. Self-organization and chaos in the metabolism of a cell. \textit{Biopolumers and Cell.} \textbf{30}, No.~5, 403 (2014).
\bibitem{31} V. Grytsay, I. Musatenko. Nonlinear self-organization dynamics of a metabolic process of the Krebs cycle. \textit{Chaotic Modeling and Simmulation (CMSIM)} \textbf{3}, 207 (2014).
\bibitem{32} V.I. Grytsay. Lupanov indices and the Poincare maping in a study of the stability of the Krebs cycle. \textit{Ukr. J. Phys.} \textbf{60}, No.~6, 561 (2015).
\bibitem{33} V.I. Grytsay. Self-oranization and fractality in the metabolic process of glycolysis. \textit{Ukr. J. Phys.} \textbf{60}, No.~12, 1251 (2015).
\bibitem{34} V. Grytsay. Self-organization and fractality created by gluconeogenesis in the metabolic process. \textit{Chaotic Model cing and Simulation (CMSIM)}, No.~2, 113 (2016).
\bibitem{35} V.I. Grytsay. Self-organization and chaos in the me\-ta\-bo\-lism of hemostasis in a blood vessel. \textit{Ukr. J. Phys.} \textbf{61}, No.~7, 648 (2016).
\bibitem{36} V.I. Grytsay. A mathematical model of the metabolic process of atheroclerosis. \textit{Ukr. Bichem. J.} \textbf{88}, No.~4, 75 (2016).
\bibitem{37} V.I. Grytsay. Spectral analysis and invariant measure in the study of a nonlinear dynamics of the metabolic process in cells. \textit{Ukr. J. Phys.} \textbf{62}, No.~5,  448
(2017).
\bibitem{38} V.I. Grytsay, A.G. Medentsev, A.Yu. Arinbasarova. Autoocillatory dynamics in mathematical model of the metabolic process in aerobic bacteria. Influence of the Krebs cycle on the self-organization of a biosytem. \textit{Ukr. J. Phys.} \textbf{65}, No. 5, 393
(2020).
\bibitem{39} V.I. Grytsay. Spectral analysis and invariant measure in studies of the dynamics of the hemostasis of a blood vessel. \textit{Ukr J. Phys.}  \textbf{66}, No.~3, 221 (2021).
\bibitem{40} V. Grytsay. Spectral analysis and invariant measure in studies of the dynamics of the Krebs cycle. \textit{Chaotic Modeling and Simulation (CMSIM)} No.~1, 35
(2021).
\bibitem{41} E. Buzaneva, A. Karlash, K. Yakovkin, Ya. Shtogun, S.~Pu\-tselyk, D. Zherebetskiy, A. Gorchinskiy, G. Popova, S.~Pri\-lutska, O. Matyshevska, Yu.I. Prylutskyy, P. Lytvyn, P.~Scharff, P. Eklund. DNA nanotechnology of carbon na\-no\-tube cells: Physico-chemical models of self-or\-ga\-ni\-za\-tion and properties. \textit{Mater. Sci. Engineer. C} \textbf{19}, Nos.~1--2, 41 (2002).
\bibitem{42} M.I. Melnyk, I.V. Ivanova, D.O. Dryn, Yu.I. Prylutskyy, V.V. Hurmach, M. Platonov, L.T. Al Kury, U. Ritter, A.I.~So\-loviev, A.V. Zholos. C$_{60}$ fullerenes selectively inhibit BK$_{\rm Ca}$ but not $\rm K_v$ channels in pulmonary artery smooth muscle cells. \textit{Nanotechnology, Biology and Medicine} \textbf{19}, 1 (2019).
\bibitem{43} L.T. Al Kury, D. Papandreou, V.V. Hurmach, D.O. Dryn, M.I. Melnyk, M.O. Platonov, Yu.I. Prylutskyy, U. Ritter, P. Scharff, A.V. Zholos. Single-walled carbon nanotubes inhibit TRPC4-mediated muscarinic cation current in mouse ileal myocytes. \textit{Nanomater.} \textbf{11}, No.~12: 3410 (2021).\vspace*{1mm}
\begin{flushright}
{\footnotesize Received 03.08.22}
\end{flushright}
\end{thebibliography}
\end{document}